\documentclass[ALICE,manyauthors]{cernphprep}

\usepackage[comma,square,numbers,sort&compress]{natbib}
\usepackage{hyperref}
\usepackage{lineno}
\usepackage{subfigure}
\usepackage[section]{placeins}
\usepackage {multicol}
\usepackage {multirow}
\usepackage {units}
\usepackage {array}
\usepackage {xcolor}
\usepackage{comment}
\usepackage{hyphenat}
\usepackage[T1]{fontenc}
\usepackage{orcidlink}

\newcommand{\fzero}{$\mathrm{f}_{0} (980)$}
\newcommand{\kstar}{$\mathrm{K}^{*}$(892)$^{0}$}
\newcommand{\rhoz}{$\rho$(770)$^{0}$}

\begin{document}

\begin{titlepage}
\PHyear{2025}
\PHnumber{146}      
\PHdate{01 July}  

%

\title{Multiplicity dependence of \textbf{f$_{\mathbf{0}}$}(980) production in pp collisions at $\mathbf{\sqrt{s}}=$~13~TeV}
\ShortTitle{}   

\Collaboration{ALICE Collaboration\thanks{See Appendix~\ref{app:collab} for the list of collaboration members}}
\ShortAuthor{ALICE Collaboration} 

\begin{abstract}
The dependence of $\mathrm{f}_{0}$(980) production on the final-state charged-particle multiplicity is reported for proton--proton (pp) collisions at the centre-of-mass energy, $\sqrt{s}=$~13~TeV. The production of $\mathrm{f}_{0}$(980) is measured with the ALICE detector via the $\mathrm{f}_0 (980) \rightarrow \pi^{+}\pi^{-}$ decay channel in a midrapidity region of $|y|<$~0.5. The evolution of the integrated yields and mean transverse momentum of f$_{0}$(980) as a function of charged-particle multiplicity measured in pp at $\sqrt{s}=$~13~TeV follows the trends observed in pp at $\sqrt{s}=$~5.02~TeV and in proton--lead (p--Pb) collisions at $\sqrt{s_{\rm{NN}}}=$~5.02~TeV. Particle yield ratios of $\mathrm{f}_{0}$(980) to $\pi^{\pm}$ and $\mathrm{K}^{*}$(892)$^{0}$ are found to decrease with increasing charged-particle multiplicity. These particle ratios are compared with calculations from the canonical statistical thermal model as a function of charged-particle multiplicity. The thermal model calculations provide a better description of the
decreasing trend of particle ratios when no strange or antistrange quark composition for f$_{0}$(980) is assumed, which suggests that the data do not support significant hidden strangeness in the \fzero.
\end{abstract}
 
\end{titlepage}

\setcounter{page}{2}

\section{Introduction}
\label{sec:intro}

Quantum chromodynamics (QCD) is the theory of the strong interaction. It describes the interactions between coloured quarks and gluons, and the formation of colour-neutral hadronic states.
In addition to the conventional meson (q$\bar{\mathrm{q}}$) and baryon (qqq) states described by the quark model~\cite{GellMann:1964ewy}, exotic states such as tetraquarks~\cite{Liu:2019zoy, Agaev:2022pis, ALICE:2021ovd}, pentaquarks~\cite{Liu:2019zoy, LHCb:2015yax}, hexaquarks~\cite{ALICE:2020mfd}, and other compositions of a higher number of constituent quarks~\cite{Park:2017mdp} are allowed, which are of interest
in theoretical and experimental research~\cite{ExHIC:2017smd}.

Light scalar mesons, one of the most frequently revisited hadron classes, consist of light-flavoured quarks and are characterized by quantum numbers $I^{G}(J^{PC}) = 0^{+}(0^{++})$ and masses below 2~GeV/$c^{2}$.
The study of light scalar mesons, including their internal structure, properties such as decay width, and formation mechanisms, is an actively explored research area~\cite{Jaffe:1976ig, Jaffe:1976ih, Amsler:2004ps}.
The \fzero~is a light scalar meson and has a large mass ($m \approx 980$~MeV/$c^{2}$) and a relatively small particle width ($\Gamma \approx$~40--70~MeV/$c^{2}$)~\cite{ParticleDataGroup:2022pth,ALICE:2023cxn}. 
Given these characteristics, the \fzero~can currently be experimentally identified compared to other light scalar states, such as $\mathrm{f}_{0} (500)$ and $\mathrm{a}_{0} (980)$, where the threshold mass lies close to the pole mass. 
Theoretical interpretations of the elusive internal structure of this particle suggest that \fzero~can either be a conventional meson ($\rm{}q\bar{q}$)~\cite{Chen:2003za}, a compact tetraquark~\cite{Achasov:2020aun}, or a $\rm{}K\overline{K}$ molecule~\cite{Ahmed:2020kmp}. 
These competing scenarios remain the subject of intense theoretical debate~\cite{ParticleDataGroup:2022pth}.

QCD predicts the formation of a deconfined and strongly interacting matter, called the quark--gluon plasma (QGP), under conditions of high temperature and high energy density~\cite{Adams:2005dq, Adcox:2004mh, Heinz:2000bk, ALICE:2022wpn}. As the system cools down, the near-perfect fluidity of QGP collapses, confining quarks and gluons into hadrons upon crossing its critical temperature. 
As the system cools below the critical temperature, all inelastic interactions stop, confining partons into hadrons, where the relative particle composition is expected to be fixed at the chemical freeze-out stage. 
In heavy-ion collisions, the thermally produced hadron gas can be modelled according to a grand canonical ensemble of particle species, considering the conservation of baryon number, electric charge, strangeness, and other quantum numbers on average. 
Statistical hadronisation modelling of particle production can be extended to pp and p--Pb collisions by employing a canonical approach. An implementation of the statistical model is the canonical statistical model with undersaturation of strangeness, $\gamma_{S}$CSM~\cite{Vovchenko:2019kes}. The $\gamma_{S}$CSM accounts for the system-size-dependent hadrochemistry at vanishing baryon density with local conservation of electric charge, baryon density, as well as strangeness. It can, therefore, predict the relative particle composition while suppressing the production of strangeness in small collision systems. Although $\gamma_{S}$CSM does not account for the hadronic phase or the rescattering of decay products from short-lived resonances, it remains a powerful tool for investigating strangeness production. In particular, such an approach can provide insight into whether certain hadrons include strange quarks.
Due to ongoing studies on the possible exotic nature of \fzero~and its production mechanisms, it has not yet been implemented in most QCD-inspired general-purpose Monte Carlo models utilised in the field.
Therefore, the $\gamma_{S}$CSM offers a suitable framework for studying the strangeness-related properties of the \fzero.

In this context, two limiting hypotheses for the strangeness content of the \fzero~are considered within the $\gamma_{S}$CSM. The first case, $|S| = 0$, corresponds to a state without valence strange quarks, consistent with light-quark configurations such as mixtures of $\rm u\bar{u}$ and $\rm d\bar{d}$ components. The second case, $|S| = 2$, represents scenarios in which the \fzero~carries hidden strangeness through the presence of an $\rm s\bar{s}$ pair. This assignment covers different possible internal structures, including $\rm s\bar{s}$-rich quark–antiquark states, four-quark configurations, or $\rm K\overline{K}$ molecular components. Here, $|S| = 2$ does not specify a particular internal arrangement, but simply encodes the presence of hidden strangeness that determines the model prediction.
For instance, the enhancement of \fzero$/\pi$~ratio with increasing multiplicity can indicate the presence of strange quarks inside the \fzero. On the other hand, the absence of late hadronic interactions in the statistical hadronisation model results in the overestimation of resonance yields, especially in large collision systems. Therefore, the model overestimates the observed $\mathrm{f}_{0}$(980)/$\pi$~ratio.
One open point, which is relevant to a precise data-to-model comparison~\cite{Reichert:2024stb}, is the knowledge of the $\mathrm{f}_{0}$(980) branching ratios (B.R.). Only few measurements of the B.R. of the $\mathrm{f}_{0}(980) \rightarrow \pi^{+}\pi^{-}$ decay channel~\cite{CLEO:2009ugx, BaBar:2006hyf, Stone:2013eaa} are available. The CLEO~\cite{CLEO:2009ugx} and BaBar~\cite{BaBar:2006hyf} measurements of the ratio of B.R. are compatible with zero within 3 units of their uncertainties, where a ratio consistent with zero indicates the absence of the $\mathrm{f}_{0}(980)\rightarrow \mathrm{KK}$ decay channel. Such a measurement of the B.R. is relevant because it directly affects the corrected yields.
However, after chemical freeze-out, the system still evolves through a hadronic phase, during which elastic interactions and resonance dynamics play an important role.

In the hadronic phase, the system is a dense gas of hadrons and resonances.
In particular, strongly decaying states have short lifetimes ($\tau\approx$~1--10$~\mathrm{fm/}c$), often comparable to the duration of the hadronic phase ($t\approx$~1--10$~\mathrm{fm/}c$).
As a result, these particles can decay within the medium, and their decay products may undergo rescattering~\cite{Steinheimer:2017vju} with surrounding hadrons.
At the same time, resonance regeneration~\cite{Steinheimer:2017vju} can occur from interactions between particle pairs in the hadronic phase. 
In this paper, we refer to these two processes collectively as hadronic interactions, which are believed to modify the measurable yields of resonances.
Short-lived resonances, such as \rhoz~\cite{ALICE:2018qdv}, \kstar~\cite{ALICE:2019etb, ALICE:2016sak, ALICE:2023edr}, K$^{*}$(892)$^{\pm}$~\cite{ALICE:2023ifn},  $\Sigma(1385)^{\pm}$~\cite{ALICE:2022zuc, ALICE:2023egx}, $\Lambda$(1520)~\cite{ALICE:2018ewo}, and \fzero~\cite{ALICE:2023cxn}, serve as important probes of the late hadronic phase.
Their yields can be compared to those of long-lived or ground-state particles to study the effects of hadronic interactions.
The ratios of $\rho^0/(\pi^{+}+\pi^{-})$ and $\mathrm{K}^{*}/(\mathrm{K}^{+}+\mathrm{K}^{-})$ are particularly illustrative. 
Notably, extracting ratios of particles with the same strangeness content mitigates potential strangeness enhancement effects, thus permitting direct interpretation of these modifications. 
Recently, system-size dependent modifications of particle yields have been observed in high-multiplicity pp and p--Pb collisions and have been interpreted as a result of hadronic interaction effects that occur in these small collision systems~\cite{ALICE:2016sak, ALICE:2019etb}. 
The measurement of the modification of the \fzero~yield may contribute to further understanding of the late hadronic phase owing to its short lifetime ($\tau\approx$~5--10$~\mathrm{fm/}c$)~\cite{ParticleDataGroup:2022pth}.

This paper reports measurements of \fzero~production in pp collisions at the centre-of-mass energy of $\sqrt{s}=$~13~TeV. The \fzero~has been measured at midrapidity ($|y|<$~0.5) in the transverse momentum ($p_{\mathrm{T}}$) range of 0~$<p_{\mathrm{T}}<$~13~GeV/$c$ and for different final-state charged-particle multiplicity classes. Section~\ref{sec:setup} describes the experimental setup. In Sec.~\ref{sec:ana}, the reconstruction of \fzero\ and the relative corrections are detailed. The systematic uncertainties in the measurement are described in Sec.~\ref{sec:syst}. Section~\ref{sec:results} presents the results, including $p_{\mathrm{T}}$ spectra, $p_{\mathrm{T}}$-integrated yields, mean $p_{\mathrm{T}}$, and the particle yield ratios. Conclusions are drawn in Sec.~\ref{sec:summary}.

\section{Experimental setup}
\label{sec:setup}
The data sample used in the analysis was collected with the ALICE detector during the LHC Run~2 (2015--2018) of pp collisions at $\sqrt{s}$~=~13~TeV with a minimum bias (MB) trigger.
Details of the ALICE apparatus during the LHC Run~2 are described in Refs~\cite{Abelev:2014ffa,ALICE:2022wpn}.
The following detectors are used in the analysis: the V0~\cite{Cortese:2004aa,ALICE:2013axi}, the Inner Tracking System (ITS)~\cite{ALICE:2010tia}, the Time Projection Chamber (TPC)~\cite{Alme:2010ke}, and the Time-Of-Flight (TOF)~\cite{Akindinov:2013tea}. 

The V0 detector consists of two arrays (V0A and V0C) of 32 scintillators located on both sides of the interaction point, covering the full azimuthal angle within the pseudorapidity intervals $2.8 < \eta < 5.1$ and $-3.7 < \eta < -1.7$, respectively. 
The MB pp collisions are selected online by requiring a signal in both V0A and V0C arrays in coincidence with a LHC bunch crossing. 
The total charge deposited in the V0A and V0C, which is called V0M, is utilised to determine the multiplicity classes. 
The multiplicity classes are defined based on the V0M amplitudes, which are proportional to the multiplicity of charged particles in the forward and backward rapidity regions. 
The event class of INEL~$>$~0 is defined as at least one charged track at midrapidity. 
The MB sample utilised in the present analysis corresponds to an integrated luminosity of $\mathcal{L}_\mathrm{int}$~=~\unit[19] {$\mathrm{nb}^{-1}$}~\cite{ALICE-PUBLIC-2021-005}.

The ITS is composed of six cylindrical layers of silicon detectors, which are arranged in the radial direction from the beam pipe. 
The innermost two layers of the ITS, the Silicon Pixel Detector (SPD)~\cite{Santoro2009:ALICESPD}, provide high-precision points in the vicinity of the primary vertex (PV) of the collision.
The PV position along the beam direction ($z_\mathrm{vtx}$) is required to be in $|z_\mathrm{vtx}|<10$~cm from the nominal interaction point ($z_\mathrm{vtx}=$~$0$). 
Pileup events are filtered by rejecting events with multiple reconstructed vertices with the additional requirement that the distance between the PV and any additional reconstructed vertex is larger than $0.8$~cm. 
In addition, a mismatch between the number of track candidates in the ITS and the clusters in the SPD is utilised to further suppress the pileup events~\cite{ALICE:2015olq}. 
After these selections, the probability of pileup events is expected to be about 0.1\% in the MB sample~\cite{ALICE:2017svf}. 
Charged particles are reconstructed down to $p_{\mathrm{T}}=$~$0.15$~GeV/$c$ in the pseudorapidity range $|\eta|<$~$0.8$ over the full azimuth with the TPC and the ITS detectors. 
These detectors are located inside a large solenoidal magnet, providing a uniform magnetic field of $0.5$~T directed along the beam axis. 
Particle identification (PID) is performed with the TPC and TOF. 
The TPC measures specific ionisation energy loss $\mathrm{d}E/\mathrm{d}x$ of charged tracks to identify particle species. 
The TOF is used for additional PID by measuring the flight time of charged particles from the PV to the TOF.

\section{Data analysis}

The \fzero~resonance is reconstructed via the decay channel \fzero~$\rightarrow \pi^{+}\pi^{-}$, with the reported B.R. of ($46\pm6$)\%~\cite{Stone:2013eaa}. The \fzero~candidates are identified through pairs of charged tracks reconstructed using the ITS and TPC. The selection criteria for the reconstructed tracks include $p_{\mathrm{T}}>$~0.15~GeV/$c$ and $|\eta|<$~0.8, to ensure uniform detector acceptance.  Further details on the quality of selection criteria used in this analysis are available in~\cite{ALICE:2020nkc}. To guarantee good momentum resolution, the tracks must traverse at least 70 readout pad rows in the TPC  (out of a maximum of 159) and have a minimum of two hits in the ITS (out of a maximum of 6), including at least one in the SPD.
To minimise the contamination from secondary charged particles originating from weakly decaying hadrons or interactions with the detector material, selection criteria are applied to the distance of closest approach (DCA) to the PV. 
The transverse component, $\mathrm{DCA}_{xy}$, is subject to a $p_{\mathrm{T}}$-dependent cut, while the longitudinal component, $\mathrm{DCA}_{z}$, has a fixed threshold. 
Specifically, $\mathrm{DCA}_{z}$ must be less than 2~cm, and $\mathrm{DCA}_{xy}$ must be less than $(0.0105 + 0.0350 \times p_{\mathrm{T}}^{-1.1})$~cm, where $p_{\mathrm{T}}$ is in GeV/$c$.

The identification of charged pions is performed by using the combined information of the TPC and TOF~\cite{ALICE:2023yuk}. Inside the TPC, the ionisation energy loss is required to be within two standard deviations of the expected value from a Bethe–Bloch parameterisation for pions. In the TOF, the measured flight time should also fall within three standard deviations of the expected flight time for a pion. If the TOF signal is unavailable, pion identification relies solely on the energy loss information provided by the TPC.

\label{sec:ana}
\begin{figure}[hbt!]
	\centering
	\subfigure{ \includegraphics[width=0.47 \textwidth]{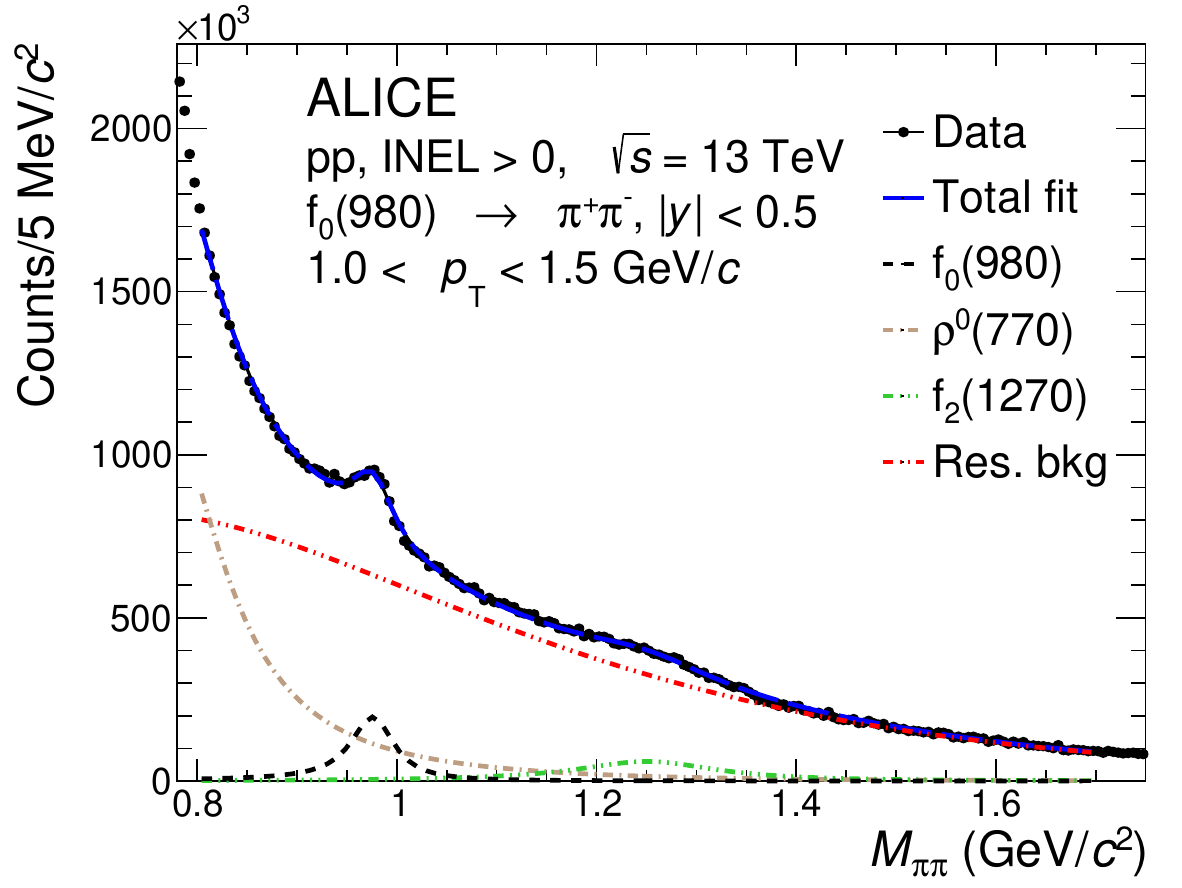} }
	\subfigure{ \includegraphics[width=0.47 \textwidth]{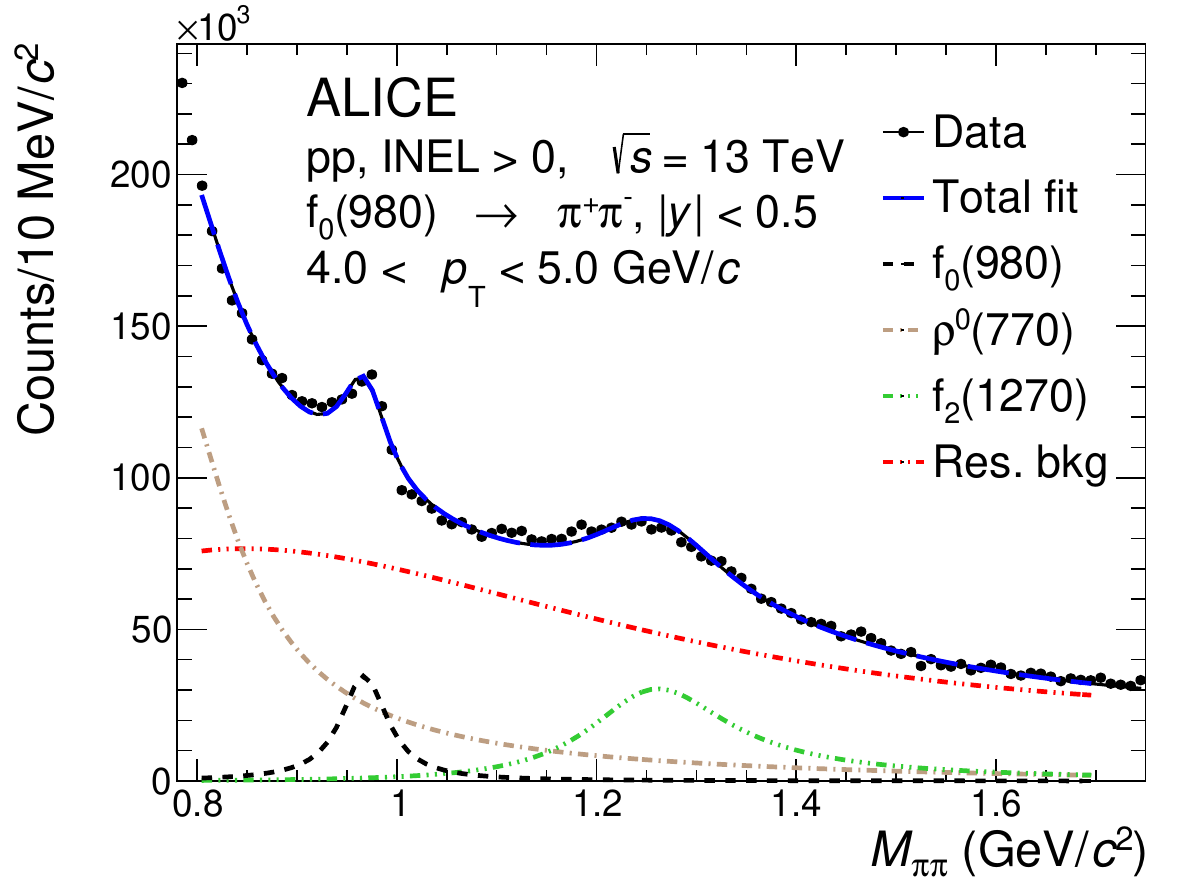} }
	\caption{ Invariant mass distributions of $\pi^{+}\pi^{-}$ pairs in $|y|<$~0.5 after the like-sign background subtraction in MB pp collisions at $\sqrt{s}=$~13~TeV. The left (right) plot is obtained using low (high) $p_{\mathrm{T}}$ of $\pi^{+}\pi^{-}$ pairs.}
	\label{fig:SigExt}
\end{figure}

The \fzero~signal is extracted through an invariant mass analysis by pairing two oppositely charged pions from the same event within the pair rapidity range $|y|<$~0.5.
The combinatorial background, arising from random combinations of uncorrelated pion pairs, is subtracted using the like-sign method~\cite{ALICE:2018qdv,LIKESIGN}. 
The like-sign background is generated by taking the geometric average of $\pi^{+}\pi^{+}$ and $\pi^{-}\pi^{-}$ distributions  from the same events, which is equal to 2$\sqrt{N_{\pi^{+}\pi^{+}}N_{\pi^{-}\pi^{-}}}$.
After subtracting the like-sign background from the signal or unlike pairs $\pi^{+}\pi^{-}$ distribution, the resulting distribution reveals peaks indicative of resonances decaying to $\pi^{+}\pi^{-}$.
Figure~\ref{fig:SigExt} shows the $\pi^{+}\pi^{-}$ like-sign-subtracted invariant mass ($M_{\pi\pi}$) distributions for 1.0~$<p_{\rm{T}}<$~1.5~GeV/$c$ (4.0~$<p_{\rm{T}}<$~5.0~GeV/$c$) in MB events on the left (right) panel.

Since \rhoz~and $\rm{f}_{2}$(1270) both decay dominantly to $\pi^{+}\pi^{-}$ and have broad widths, the \fzero~candidates include contributions from these two resonances.
Furthermore, a residual background ($f_{\mathrm{bkg}}$), which is mainly arising from misidentified particles and mini-jets, is present.
To disentangle the contributions of these backgrounds and the three resonances, the measured invariant mass distribution is modelled with a superposition of several functions.
Each resonance contribution is described with a relativistic Breit-Wigner function (rBW)~\cite{ALICE:2018qdv, ALICE:2022qnb}.
It is important to note that the detector resolution of $\mathcal{O}$(1 MeV) gives a negligible contribution to the widths of these broad resonances~\cite{ALICE:2016sak}.
The rBW can be expressed as
\begin{eqnarray}
\mathrm{rBW}(M_{\pi\pi}) = \dfrac{AM_{\pi\pi}\Gamma(M_{\pi\pi})M_{0}}{(M_{\pi\pi}^{2}-M_{0}^{2})^{2} + M_{0}^{2}\Gamma^{2}(M_{\pi\pi})},
\label{eq:rBW}
\end{eqnarray}

where $\Gamma(M_{\pi\pi})$ is the relativistic particle width and can be expressed as
\begin{eqnarray}
\Gamma(M_{\pi\pi}) = \left[ \dfrac{ (M_{\pi\pi}^{2} - 4m_{\pi}^{2}) }{ (M_{0}^{2}-4m_{\pi}^{2}) } \right]^{(2J+1)/2} \times \dfrac{\Gamma_{0}M_{0}}{M_{\pi\pi}} .
\label{eq:rBWW}
\end{eqnarray}

Here, $A$, $M_{0}$, and $\Gamma_{0}$ denote the amplitude, the rest mass, and the width of the resonance, respectively.
The spin of the resonance is represented as $J$, and the charged pion mass ($m_{\pi}$), is 139.6~MeV/$c^{2}$~\cite{ParticleDataGroup:2022pth}.
The spins for the \fzero, \rhoz, and $\mathrm{f}_{2}$(1270) resonances are 0, 1, and 2, respectively. In order to correct for the phase space effects, each resonance rBW is multiplied by the phase space factor ($\mathrm{PS}(M_{\pi\pi})$)~\cite{ALICE:2018qdv}, which can be expressed as
\begin{eqnarray}
\mathrm{PS}(M_{\pi\pi}) = \dfrac{M_{\pi\pi}}{\sqrt{M_{\pi\pi}^{2}+p_{\mathrm{T}}^{2}}}\times\exp{\left(-\sqrt{M_{\pi\pi}^{2}+p_{\mathrm{T}}^{2}}/T_{\mathrm{kin}}\right)},
\label{eq:ps}
\end{eqnarray}

where $p_{\mathrm{T}}$ represents the transverse momentum of the $\pi\pi$ pair and is set to the median for each $p_{\mathrm{T}}$ interval, and $T_{\mathrm{kin}}$, the kinetic freeze-out temperature, is set to 160~MeV for all defined multiplicity classes~\cite{ALICE:2018qdv}, and is varied to estimate the corresponding systematic uncertainties. The $T_{\mathrm{kin}}$ can be determined based on a Blast-wave analysis and is observed to vary with the multiplicity class~\cite{ALICE:2020nkc}. The fit function for the residual background ($f_{\mathrm{bkg}}$) is modelled using a Maxwell-Boltzmann-like distribution~\cite{OPAL:1998enc}, which can be expressed as
\begin{eqnarray}
f_{\mathrm{bkg}}(M_{\pi\pi}) = B(M_{\pi\pi}-2m_{\pi})^{n}\exp{(c_{1}M_{\pi\pi} + c_{2}M_{\pi\pi}^{2})},
\label{eq:bkg}
\end{eqnarray} 

where, $B$, $n$, $c_{1}$, and $c_{2}$ are free parameters. 

The total fit function is composed of three resonance rBWs and the $f_{\mathrm{bkg}}$.
The fit function has nine free parameters, including three for the \fzero~resonance (mass, width, and amplitude), two amplitudes for the \rhoz~and f$_{2}$(1270) resonances, and four parameters for $f_{\mathrm{bkg}}$.
The width of the \fzero~resonance is not yet precisely determined by measurements and is, therefore, left as a free parameter in the fit, with a range of 10~$<\Gamma_{0}^{\mathrm{f}_{0}}<$~100~MeV/$c^{2}$~\cite{ParticleDataGroup:2022pth}.
The masses and widths of \rhoz~and $\mathrm{f}_{2}$(1270) are fixed to their world-average values from Ref.~\cite{ParticleDataGroup:2022pth}, namely $m_{\rho}=$~766.5~MeV/$c^{2}$, $\Gamma^{\rho}_{0}=$~~149.1~MeV/$c^{2}$, $m_{\mathrm{f}_{2}}=$~1,275.5~MeV/$c^{2}$, and $\Gamma^{\mathrm{f}_{2}}_{0}=$~186.7~MeV/$c^{2}$. 

Due to the large number of free parameters in the fit function, the fitting procedure is divided into three distinct steps, designed to prevent the convergence of parameter values at local minima.
At the beginning, an initial value for the \fzero~width is determined using the MB sample, employing the same $p_{\mathrm{T}}$ binning to minimise the impact of statistical fluctuations. 
At this stage, all nine parameters are allowed to vary freely. 
The subsequent step focuses on refining the parameters of $f_{\mathrm{bkg}}$, while the \fzero~width is fixed at the value derived from the first step.
The final step involves fixing the $f_{\mathrm{bkg}}$ parameters as determined earlier and permitting the \fzero~width to adjust within a specified range of 10~$<\Gamma_{0}^{\mathrm{f}_{0}}<$~100~MeV/$c^{2}$.
Here, the fit allows the amplitudes of the three resonances and the \fzero~mass to vary freely, with the fit range set between 0.8~$<M_{\pi\pi}<$~1.76~GeV/$c^{2}$.
Notably, the extracted width of the \fzero~resonance varies from 40 to 70~MeV/$c^{2}$ across different $p_{\mathrm{T}}$ and multiplicity intervals used in this analysis.

The raw yields of \fzero~($N_{\mathrm{f}_{0}}$) for each $p_{\mathrm{T}}$ interval are calculated by integrating the \fzero~rBW function.
These yields are subsequently corrected for the reconstruction efficiency ($\mathrm{Acc} \times \epsilon$) and  normalised for the number of selected pp collisions ($N_{\mathrm{evt}}$), the width of the $p_{\mathrm{T}}$ ($\Delta p_{\mathrm{T}}$), rapidity interval ($\Delta \mathrm{y} =$~1), and the B.R.~\cite{Stone:2013eaa}.
The fully corrected yield is given by the following equation:
\begin{eqnarray}
\dfrac{1}{N_{\mathrm{evt, corr}}}\dfrac{\mathrm{d}^{2}N}{\mathrm{d}y\mathrm{d}p_{\mathrm{T}}} = \dfrac{ \epsilon_{\mathrm{trig}} \times f_{\mathrm{vtx}}  }{N_{\mathrm{evt}}} \dfrac{ N_{\mathrm{f}_{0}} }{ \Delta \mathrm{y} \Delta p_{\mathrm{T}} } \dfrac{ f_{\mathrm{SL}} }{\mathrm{Acc} \times \epsilon \times \mathrm{B.R.} }.
\end{eqnarray}

The yield of the \fzero~resonances is corrected and normalised to the number of events in each multiplicity class, represented by $N_{\mathrm{evt, corr}}$.
This is done using the factors $\epsilon_{\mathrm{trig}} \times f_{\mathrm{vtx}}$.
The trigger efficiency $\epsilon_{\mathrm{trig}}$ and vertex reconstruction efficiency $f_{\mathrm{vtx}}$ depend on the multiplicity class, increasing from 0.85 to 1 and 0.97 to 1, respectively, as the multiplicity increases.
The coefficients for the acceptance ($\mathrm{Acc}$) and the efficiency ($\epsilon$) of the tracking and PID for pion pairs are estimated using a detailed simulation of the ALICE detector response.
The pp collisions are simulated using the PYTHIA8~\cite{Bierlich:2022pfr} event generator with the injection of \fzero~signals.
The generated signals are transported through the detector using GEANT3~\cite{Brun:1994aa}.
The $\mathrm{Acc}\times\epsilon$ is obtained from INEL$>$0 events and applied uniformly across all multiplicity classes. It ranges from 26\% at 0~$<p_{\mathrm{T}}<~$0.3~GeV/$c$ to 60\% as $p_{\mathrm{T}}$ increases, and shows no significant dependence on event multiplicity.
The $\mathrm{B.R.}$ of the \fzero~to $\pi^{+}\pi^{-}$ decay is estimated to be (46~$\pm$~6)\%~\cite{Stone:2013eaa}.
The correction for the signal loss ($f_{\mathrm{SL}}$), which compensates for signal losses due to event selection, is derived using the $\phi$ meson as a surrogate because typical Monte Carlo event generators do not generate primary \fzero~particles.
This method, exploiting universal $m_{\mathrm{T}}$ scaling~\cite{Altenkamper:2017qot}, shows that $f_{\mathrm{SL}}$ does not depend on particle species~\cite{ALICE:2019xyr}.
It is quantified at 1.08 for 0~$<p_{\mathrm{T}}<$~0.3~GeV/$c$ in 50--100\% multiplicity class and approaches unity as $p_{\mathrm{T}}$ increases. In the highest multiplicity class, $f_{\mathrm{SL}}$ is near unity for the $p_{\mathrm{T}}$ range this analysis measures.

\section{Systematic uncertainties}
\label{sec:syst}

The systematic uncertainties related to the production of \fzero~are estimated by varying the analysis selection criteria, the fit configuration utilised to extract the raw yield, and the phase space correction treatment. 
For each systematic variation, a conservative maximal requirement of 1.3 standard deviations is adopted for the Barlow test~\cite{Barlow:2002yb}. 
Table~\ref{tab:syst} summarises the estimated uncertainties. The total systematic uncertainty is calculated by adding the different contributing sources in a quadratic sum. For each source, the minimum and maximum uncertainty values are reported in the table. The relative uncertainty of the $\mathrm{B.R.}$ is 13\%~\cite{Stone:2013eaa} and is not included in the total uncertainty.

\begin{table}[h!]
\caption{Relative systematic uncertainties of the \fzero~$p_{\rm{T}}$-differential yields. Numbers given in ranges correspond to minimum and maximum uncertainties.}
\centering
\begin{tabular}{l|c}
\hline 
Sources &Systematic uncertainty (\%) \\ \hline
PV selection & negligible \\ 
Pileup rejection & negligible \\ 
Tracking & 1.4 \\
Material budget & 0.4--2 \\ 
ITS--TPC matching & 6 \\
Particle identification & 2--10 \\ 
$\mathrm{f}_{2}$(1270) parameters	& 1--7 \\ 
\rhoz~parameters & 1--8 \\
Fit range & 0--8 \\
Initial $\mathrm{f}_{0}$ width & 1--8 \\
Phase space correction & 2--8 \\ \hline 
Total (in quadrature)	& 10--18 \\ 
\hline 
\end{tabular}
\label{tab:syst}
\end{table}

The analysis is repeated with a different PV selection of $|z_\mathrm{vtx}|<$~7~cm to estimate the systematic uncertainty from the PV selection. The systematic uncertainty is found to be negligible compared to the statistical fluctuation arising from the selection variation. The systematic uncertainty from the pileup rejection is studied by varying the number of track segments used to reconstruct pileup collision vertices from 5 to 3, and the resulting uncertainty is found to be negligible.

The systematic uncertainty on the tracking is determined by varying the requirements used to select reconstructed primary tracks, such as those on $\rm{DCA}_{xy}$, $\rm{DCA}_{z}$, and the number of crossed rows. According to Ref.~\cite{ALICE:2020nkc}, this results in an estimated uncertainty of 1.4\%. The systematic uncertainty from PID is tested with different requirements on the number of standard deviations ($\pm\,0.5\sigma$ with respect to the default selections) for the TPC and TOF selection. The resulting uncertainties vary, ranging from 2\% to 10\% in the different $p_{\mathrm{T}}$ intervals and multiplicity classes.

The systematic uncertainty on the extraction of the \fzero~yield through fits to the invariant mass distributions is estimated by changing certain configurations in the fit procedure. The contributions from the masses and widths of f$_{2}$(1270) and \rhoz~are evaluated by shifting the masses and widths by three standard deviations with respect to their world-average values, using the uncertainties reported in Ref.~\cite{ParticleDataGroup:2022pth}. The estimated systematic uncertainties are 1--7\% and 1--8\%, respectively. In addition, the invariant mass range used in the fit is varied inward or outward by 40~MeV/$c^{2}$, and the resulting systematic uncertainty is found to be less than 8\%. The systematic uncertainty from the initial \fzero~width value, which is obtained in the first fit step described in Sec.~\ref{sec:ana}, is estimated by varying the width within the statistical uncertainties, which is about 5~MeV/$c^{2}$ on average, in both directions. These variations affect the background distribution determined in the second step, and the estimated systematic uncertainties range from 1\% to 8\%. The systematic uncertainty from the phase space correction is estimated by varying the kinetic freeze-out temperature in the range of 140~$<T_{\mathrm{kin}}<$~180~MeV. The estimated uncertainties range from 2\% to 8\%.

The systematic uncertainty from the ITS--TPC matching is estimated by comparing the matching efficiencies in experimental data and MC samples and is 6\%~\cite{ALICE:2020nkc}. The systematic uncertainty from the material budget is estimated by varying the amount of the budget by $\pm$5\% in the simulation and is 0.4--2\%~\cite{ALICE:2020nkc}.

The correlations between systematic uncertainties across different multiplicity classes are quantitatively assessed. A stronger correlation is expected when the systematic deviations exhibit the same direction across different multiplicity classes. To test the correlations, the directions of the systematic deviations between a given multiplicity class and the MB class are compared. The systematic uncertainties analysed in this study account for approximately half of the total systematic uncertainties, with the remaining half being uncorrelated.

\section {Results}
\label{sec:results}

\begin{figure}[!hbt]
	\centering
	\subfigure{ \includegraphics[width=0.6 \textwidth]{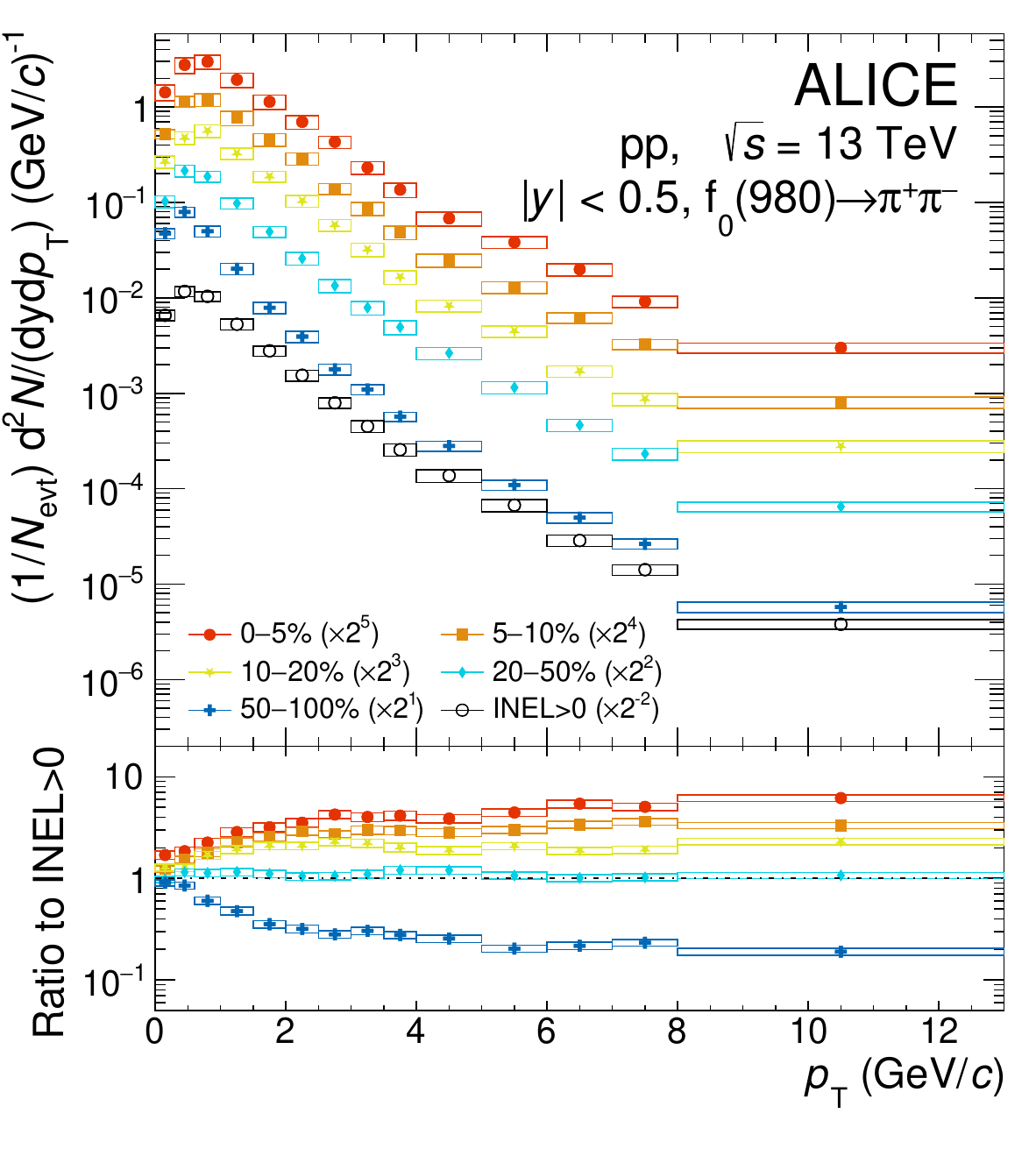} }
	\caption{ Transverse momentum spectra of \fzero~in pp collisions at $\sqrt{s}=$~13~TeV for different multiplicity classes. Each spectrum is scaled with the number shown in the figure. Statistical and systematic uncertainties are represented as bars and boxes, respectively. The normalisation uncertainty from the $\mathrm{B.R.}$ of $46\pm6$\% is not expressed in the figure. The lower panel shows the ratios of the spectra in multiplicity classes to the INEL~$>$~0 spectrum. }
	\label{fig:pt}
\end{figure}

Figure~\ref{fig:pt} shows the $p_{\mathrm{T}}$ spectra of \fzero~in pp collisions at $\sqrt{s}=$~13~TeV measured in the range of 0~$<p_{\mathrm{T}}<$~13~GeV/$c$ for different multiplicity classes and INEL~$>$~0 events, which is defined in Ref~\cite{ALICE:2020swj}. Each spectrum is scaled by a multiplicative factor, denoted in the figure, for better visibility. The lower panel of Fig.~\ref{fig:pt} shows the ratios of the $p_{\mathrm{T}}$ spectra in different multiplicity classes to the INEL~$>$~0 one. The systematic uncertainties of the ratios are calculated by propagating the multiplicity-uncorrelated uncertainties on the individual spectra. For $p_{\mathrm{T}}<$~4~GeV/$c$, a hardening of the $p_{\mathrm{T}}$ spectrum from low- to high-multiplicity events is observed, while the spectral shapes in the different multiplicity classes are found to become consistent among each other for $p_{\mathrm{T}}>$~4~GeV/$c$. Similar trends are observed also for other hadron species~\cite{Schnedermann:1993ws, ALICE:2019hno} and \fzero~measured in p--Pb collisions~\cite{ALICE:2023cxn}.  

\begin{figure}[!hbt]
	\centering
	\subfigure{\includegraphics[width=0.49\textwidth]{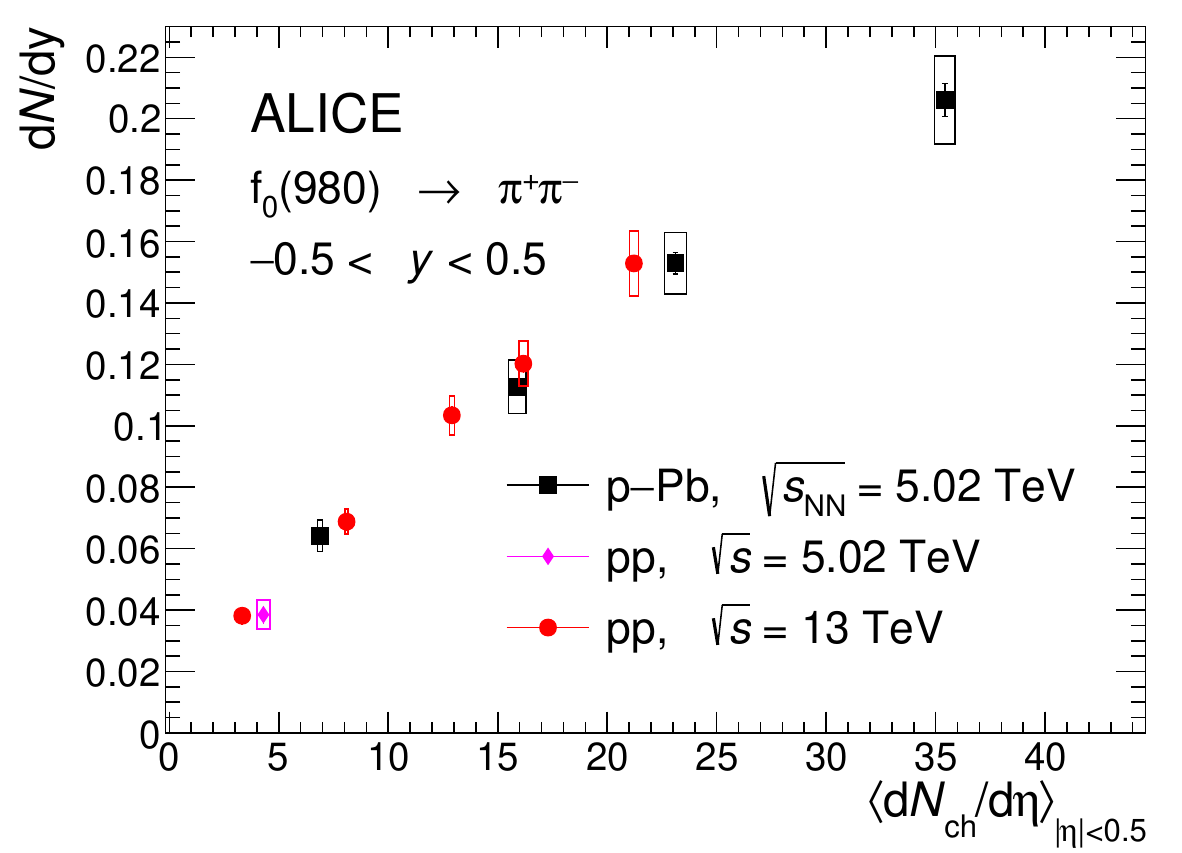} }
        \subfigure{\includegraphics[width=0.49\textwidth]{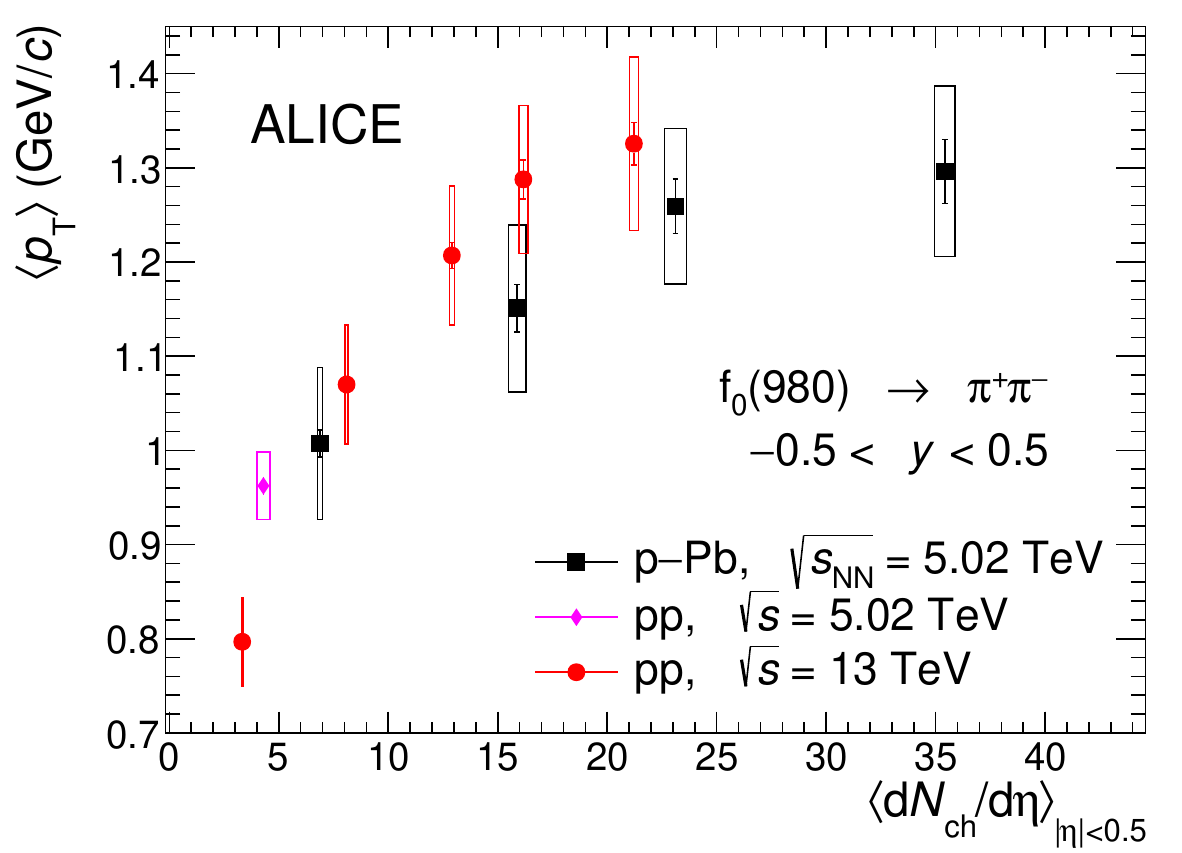} }
	\caption{ $p_{\mathrm{T}}$-integrated yield (left) and mean transverse momentum of \fzero~(right) as a function of charged-particle multiplicity. Statistical and systematic uncertainties are represented as bars and boxes, respectively. The uncertainty from the $\mathrm{B.R.}$ is not represented in the left panel. }
	\label{fig:ypt}
\end{figure}

Figure~\ref{fig:ypt} shows the $p_{\mathrm{T}}$-integrated yield (d$N/$dy) and mean transverse momentum ($\left\langle p_{\mathrm{T}} \right\rangle$) of \fzero~as a function of charged-particle multiplicity in different collision systems~\cite{ALICE:2023cxn, ALICE:2022qnb}. The d$N$/d$y$ and $\left\langle p_{\mathrm{T}} \right\rangle$ are calculated by integrating and averaging the $p_{\mathrm{T}}$ spectrum, respectively, in different multiplicity classes. The d$N/$d$y$ of \fzero~is found to increase linearly with the charged-particle multiplicity, and the $\left\langle p_{\mathrm{T}} \right\rangle$ increases with multiplicity and then seems to saturate at high multiplicity, as observed for other hadron species~\cite{ALICE:2020nkc}.
 
\begin{figure}[!hbt]
	\centering
	\includegraphics[width=0.7 \textwidth]{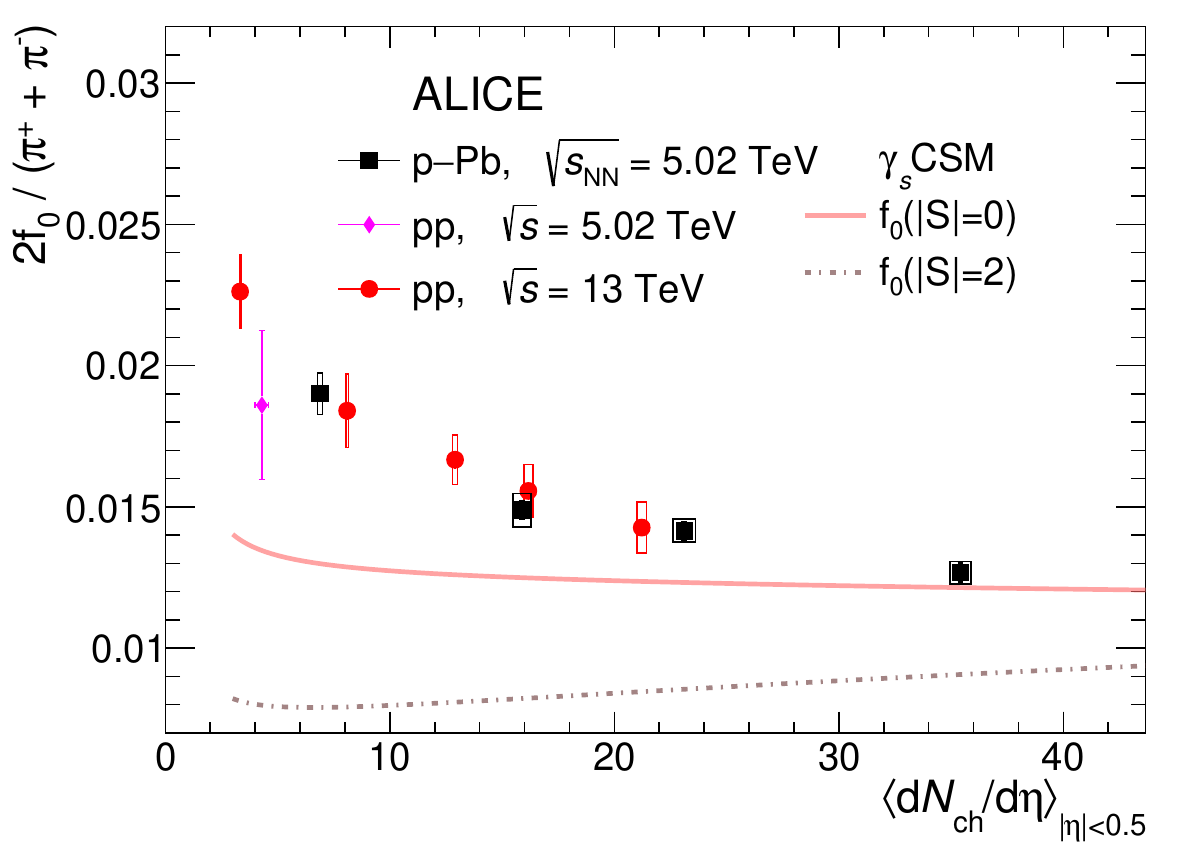} 
	\caption{ Particle yield ratio of \fzero~to charged pions as a function of charged-particle multiplicity. Vertical lines and boxes around data points represent statistical and systematic uncertainties, respectively. Two curves show calculations from $\gamma_{s}$CSM~\cite{Vovchenko:2019kes} with two assumptions for the hidden strangeness of \fzero~in the correlation volume of 3$\mathrm{d}V/\mathrm{dy}$.}
	\label{fig:f0piAddCSM}
\end{figure}

Figure~\ref{fig:f0piAddCSM} shows the particle yield ratio of \fzero~to charged pions as a function of charged-particle multiplicity in different collision systems. The \fzero$/\pi$ decreases with increasing multiplicity across different collision systems. The experimental data are compared with theoretical calculations from the $\gamma_{S}$CSM~\cite{Vovchenko:2019kes}. The $\gamma_{S}$CSM calculation with the assumption of $|S|=$~2 for \fzero~expects \fzero$/\pi$ to increase with increasing multiplicity because it assumes a specific quark content for the \fzero, including strange quarks. 
On the other hand, the $\gamma_{S}$CSM calculation with the assumption of $|S|=$~0 for \fzero~qualitatively describes the decreasing trend, which is observed in experimental data. However, a quantitative description is not achieved as the model is consistent only with the point in the highest-multiplicity events, and the model provides an underestimation of 39\% of the particle yield ratio in the lowest-multiplicity events. The model provides good description of particle yield ratio in low-multiplicity events and overestimates the ratio in high-multiplicity events for other hadron species~\cite{Vovchenko:2019kes}.
In both the $|S|=2$ and $|S|=0$ scenarios, an additional decrease with increasing multiplicity can be expected if the rescattering effect is considered, and in the $|S|=2$ scenario, the increase could be converted to a decrease. The model prediction for the $|S|=0$ scenario is closer to the data.

\begin{figure}[!hbt]
	\centering
	\includegraphics[width=0.7 \textwidth]{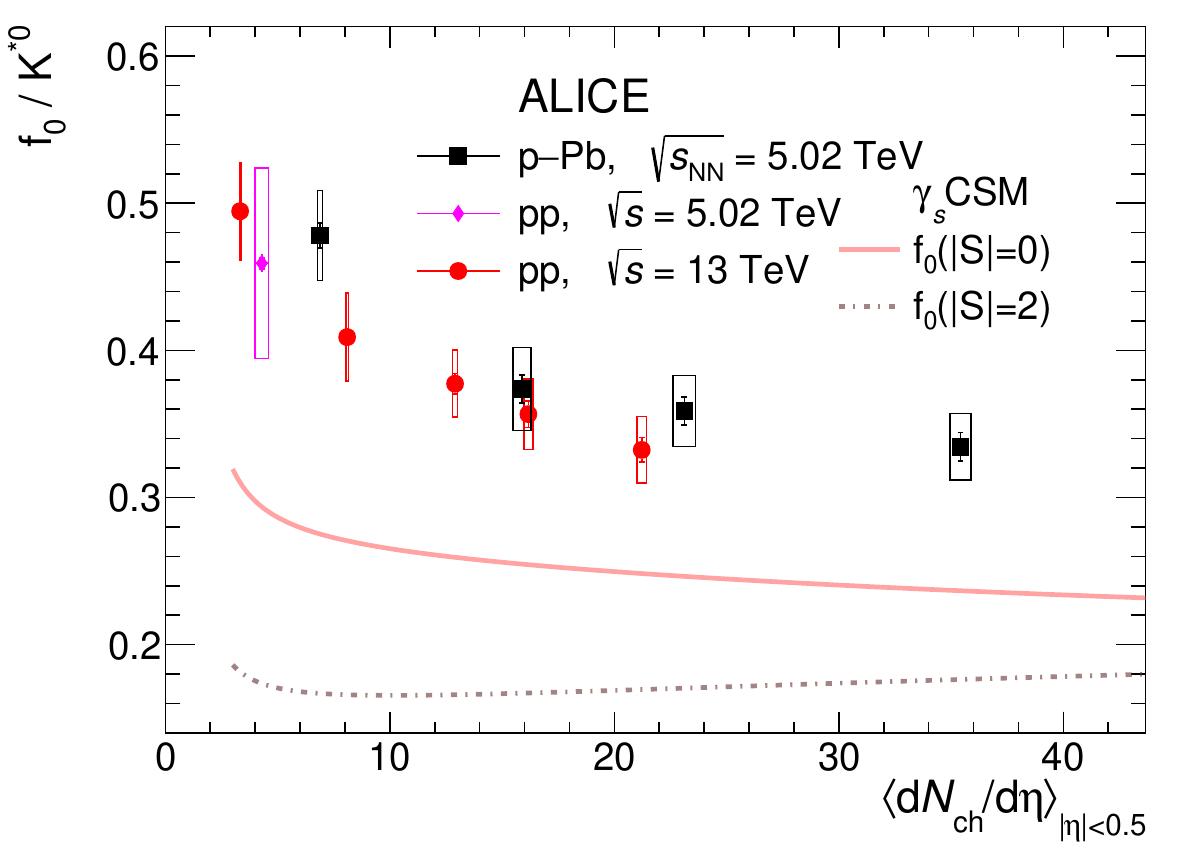} 
	\caption{ Particle yield ratio of \fzero~to K$^{*}$(892)$^{0}$ as a function of charged-particle multiplicity. Vertical lines and boxes around data points represent statistical and systematic uncertainties, respectively. Two curves show calculations from $\gamma_{S}$CSM~\cite{Vovchenko:2019kes} with two assumptions for the hidden strangeness of \fzero~in the correlation volume of 3$\mathrm{d}V/\mathrm{dy}$.}
	\label{fig:f0ksAddCSM}
\end{figure}

Figure~\ref{fig:f0ksAddCSM} shows the particle yield ratio of \fzero~to K$^{*}$(892)$^{0}$ as a function of charged-particle multiplicity in different collision systems. The \fzero$/$K$^{*}$(892)$^{0}$ decreases with increasing multiplicity in different collision systems. The $\gamma_{S}$CSM calculation with the assumption of $|S|=$~0 also predicts the decreasing trend of the particle yield ratio, whereas the calculation with the $|S|=2$ scenario does not. However, with the $|S| = 0$ scenario, the model underestimates the particle yield ratio by about 35\% throughout the multiplicity range, which is of a similar magnitude to the underestimation observed in the \fzero/$\pi$ ratio.
Because rescattering affects both the \fzero~and \kstar~resonances, its effects are expected to partially cancel out in the \fzero/\kstar~yield ratio. 
However, despite their similar lifetimes, they differ in their decay products, where \fzero~and \kstar~decay into $\pi\pi$ and K$\pi$, respectively. The rescattering effect for \fzero~can be stronger than for \kstar~when the scattering cross section $\sigma(\pi\pi)$ is greater than $\sigma(\mathrm{K}\pi)$. Similar to \fzero$/\pi$ ratio, the model prediction for \fzero$/$\kstar~with the $|S|=0$ scenario is closer to the data.

\begin{figure}[!hbt]
	\centering
	\includegraphics[width=0.49 \textwidth]{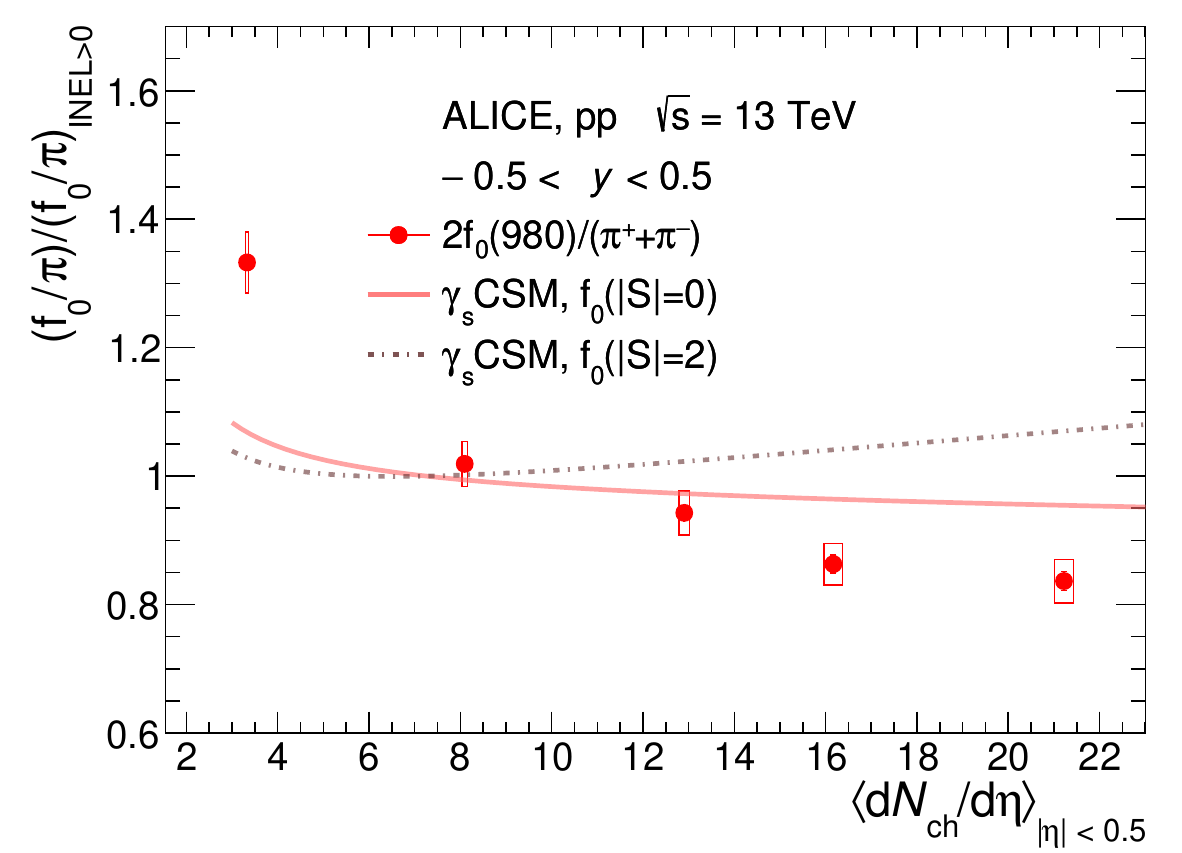} 
    \includegraphics[width=0.49 \textwidth]{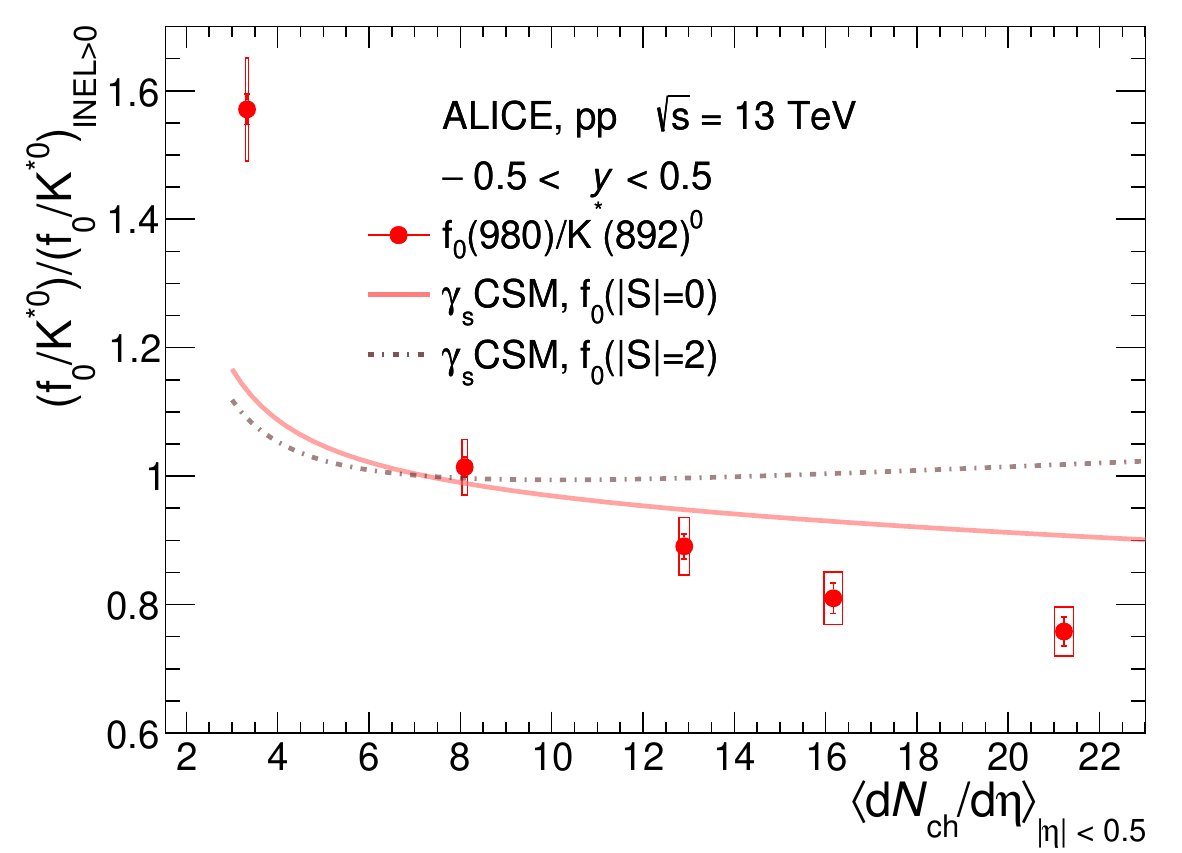} 
	\caption{Double ratios of \fzero~to $\pi$ (left panel) and K$^{*}$(892)$^{0}$ (right panel) as a function of charged-particle multiplicity. The single ratio is divided by the ratio in INEL~$>$~0 events, which cancels correlated systematic uncertainty. Two curves show calculations from $\gamma_{S}$CSM~\cite{Vovchenko:2019kes} with two assumptions for the hidden strangeness of \fzero~in the correlation volume of 3$\mathrm{d}V/\mathrm{dy}$.}
	\label{fig:ptintdr}
\end{figure}

Figure~\ref{fig:ptintdr} presents the $p_{\mathrm{T}}$-integrated particle yield double ratio of \fzero~to $\pi$ (left panel) and \kstar\hyp{}(right panel) as a function of charged-particle multiplicity in pp collisions at $\sqrt{s}=13$~TeV. The double ratio is defined as the ratio of the \fzero~to hadron yield ratio in a given multiplicity class to the corresponding value measured in INEL~$>$~0 events. This approach facilitates a direct comparison of the multiplicity dependence while mitigating correlated systematic uncertainties and cancelling uncertainties related to branching ratios. For the double ratios, the $\gamma_{S}$CSM prediction with the $|S|=0$ scenario is closer to the data.

\begin{figure}[!hbt]
	\centering
	\includegraphics[width=0.49 \textwidth]{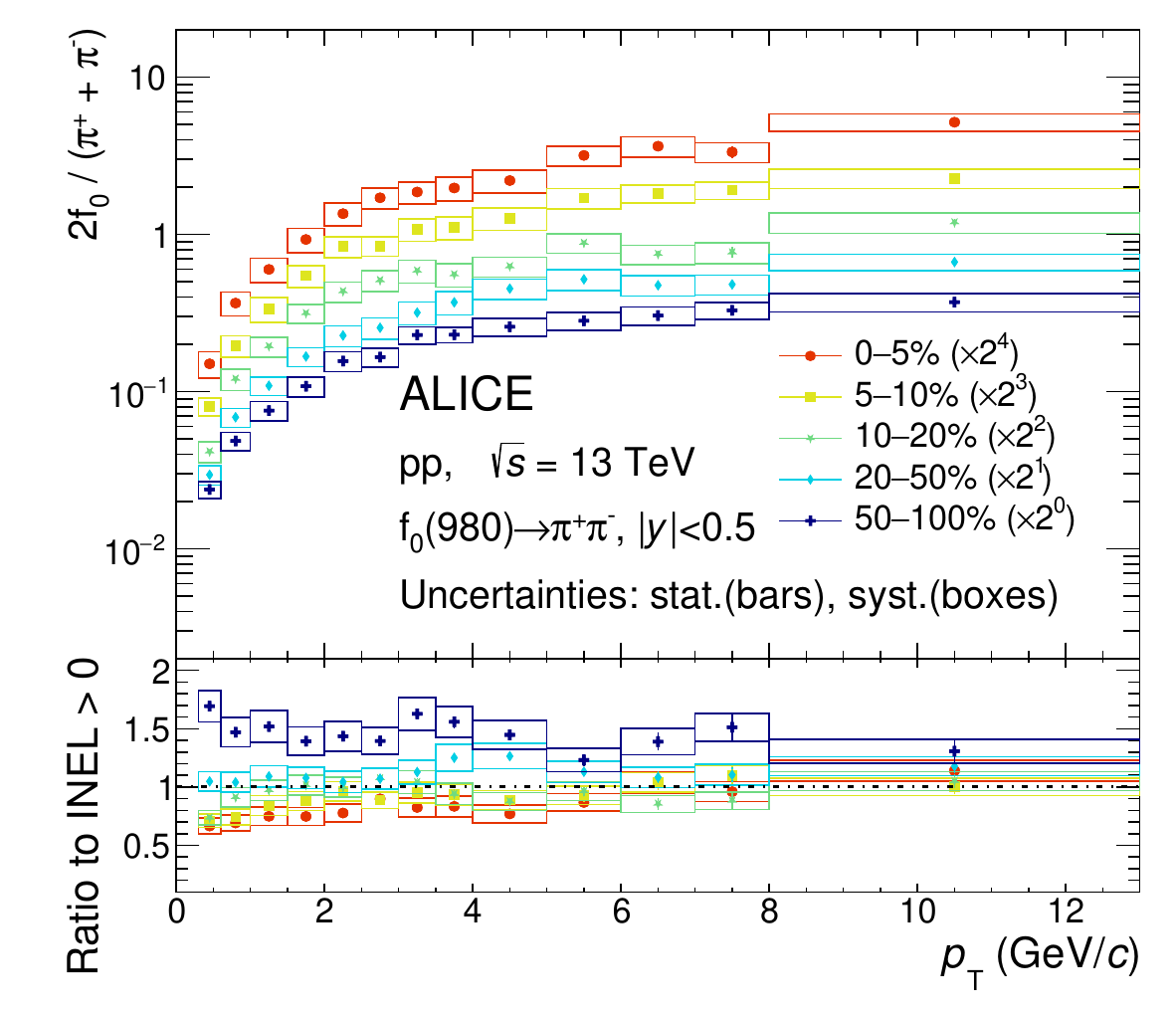} 
    \includegraphics[width=0.49 \textwidth]{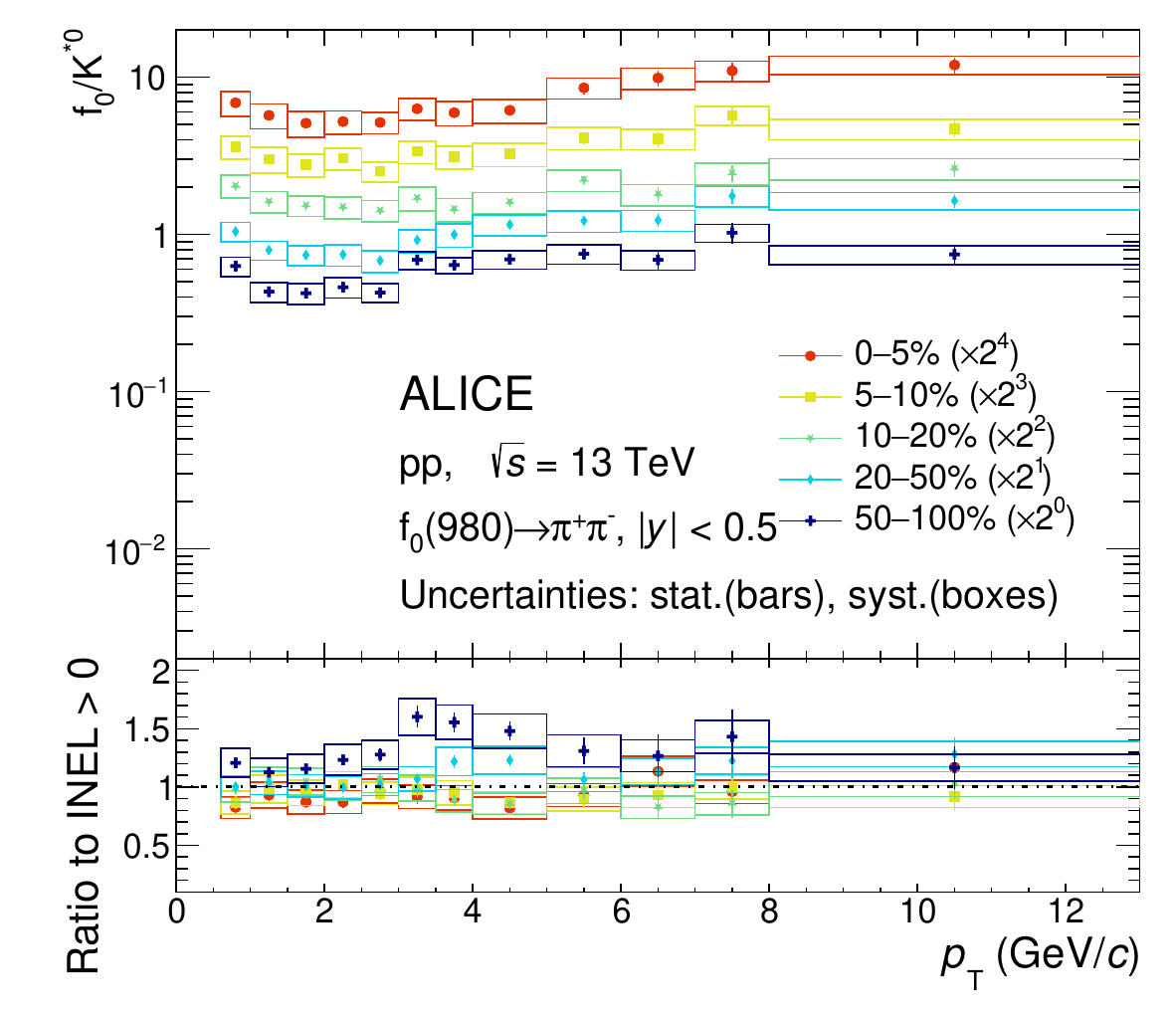} 
	\caption{Particle yield ratios of \fzero~to $\pi$ (left panel) and K$^{*}$(892)$^{0}$ (right panel) as a function of $p_{\mathrm{T}}$ for different multiplicity classes. The lower panel shows double ratios of ratio spectra to INEL~$>$~0 events, where the correlated uncertainties cancel.}
	\label{fig:ptdiffdr}
\end{figure}

Figure~\ref{fig:ptdiffdr} shows the $p_{\mathrm{T}}$-differential particle yield ratio of \fzero~to $\pi$ (left panel) and \kstar~(right panel) as a function of charged-particle multiplicity in pp collisions at $\sqrt{s}=13$~TeV, along with their double ratios relative to INEL~$>$~0 events in the lower panels. For better visibility, the single ratios are scaled by the factors indicated in the legend. The systematic uncertainties of the double ratios in the lower panels account only for uncorrelated uncertainties. The ratio of \fzero~to $\pi$ increases with $p_{\mathrm{T}}$ across all measured multiplicity classes. The double ratio at low multiplicity is higher than that at high multiplicity, particularly at low $p_{\mathrm{T}}$, and approaches unity at higher $p_{\mathrm{T}}$, suggesting that \fzero~is relatively suppressed only at low $p_{\mathrm{T}}$. 
On the other hand, the ratio of \fzero~to \kstar~is nearly constant with $p_{\mathrm{T}}$, likely due to the similar mass for both particles.
The double ratio of \fzero~to \kstar~shows a mild decrease with increasing multiplicity but exhibits no significant $p_{\mathrm{T}}$ dependence, indicating that the suppression mechanism does not vary with transverse momentum.

\section{Conclusions}
\label{sec:summary}

The multiplicity and $p_{\mathrm{T}}$ dependence of \fzero~production in pp collisions at $\sqrt{s}$~=~13~TeV is presented. The \fzero~is reconstructed via the \fzero~$\rightarrow\pi^{+}\pi^{-}$ decay channel at midrapidity ($|y|<$~0.5) in the transverse momentum region of $0<p_{\mathrm{T}}<$~13~GeV/$c$. A hardening of the $p_{\mathrm{T}}$ spectra and a consequent increase of the mean $p_{\mathrm{T}}$ are observed with increasing multiplicity.

The $p_{\mathrm{T}}$-integrated particle yield ratio of \fzero~to $\pi$ decreases with increasing multiplicity. The $\gamma_{S}$CSM calculation shows qualitatively such a trend for the $|S| = 0$ scenario
but yields to an unexpected underestimation of the ratio in low-multiplicity events. The $p_{\mathrm{T}}$-differential particle yield ratio shows that \fzero~is relatively suppressed only at low $p_{\mathrm{T}}$. These phenomena are also observed for short-lived resonances produced in heavy-ion collisions, where it is attributed to be the result of hadronic rescatterings.

The $p_{\mathrm{T}}$-integrated particle yield ratio of \fzero~to \kstar~decreases with increasing multiplicity, as expected under the $|S| = 0$ scenario. In addition, the $\gamma_{S}$CSM calculation underestimates the ratio across the measured multiplicity range. However, yields of two particles can be modified differently by hadronic interactions, which complicates the physical interpretation. This ambiguity could be reduced by measuring the particle yield ratio in large collision systems, where the strangeness enhancement is no longer present and only hadronic interaction effects remain.

The comparison between the data and model calculations suggests that the \fzero~is less likely to contain a strange or antistrange quark, and therefore is unlikely to have a tetraquark structure. Given the large uncertainty in the branching ratio of \fzero~$\rightarrow\pi^{+}\pi^{-}$ and the lack of $p_{\mathrm{T}}$ differential model calculations for hadronic interactions on the \fzero, further developments in these aspects would greatly improve our understanding of the \fzero.

\newenvironment{acknowledgement}{\relax}{\relax}
\begin{acknowledgement}
\section*{Acknowledgements}

The ALICE Collaboration would like to thank all its engineers and technicians for their invaluable contributions to the construction of the experiment and the CERN accelerator teams for the outstanding performance of the LHC complex.
The ALICE Collaboration gratefully acknowledges the resources and support provided by all Grid centres and the Worldwide LHC Computing Grid (WLCG) collaboration.
The ALICE Collaboration acknowledges the following funding agencies for their support in building and running the ALICE detector:
A. I. Alikhanyan National Science Laboratory (Yerevan Physics Institute) Foundation (ANSL), State Committee of Science and World Federation of Scientists (WFS), Armenia;
Austrian Academy of Sciences, Austrian Science Fund (FWF): [M 2467-N36] and Nationalstiftung f\"{u}r Forschung, Technologie und Entwicklung, Austria;
Ministry of Communications and High Technologies, National Nuclear Research Center, Azerbaijan;
Rede Nacional de Física de Altas Energias (Renafae), Financiadora de Estudos e Projetos (Finep), Funda\c{c}\~{a}o de Amparo \`{a} Pesquisa do Estado de S\~{a}o Paulo (FAPESP) and The Sao Paulo Research Foundation  (FAPESP), Brazil;
Bulgarian Ministry of Education and Science, within the National Roadmap for Research Infrastructures 2020-2027 (object CERN), Bulgaria;
Ministry of Education of China (MOEC) , Ministry of Science \& Technology of China (MSTC) and National Natural Science Foundation of China (NSFC), China;
Ministry of Science and Education and Croatian Science Foundation, Croatia;
Centro de Aplicaciones Tecnol\'{o}gicas y Desarrollo Nuclear (CEADEN), Cubaenerg\'{\i}a, Cuba;
Ministry of Education, Youth and Sports of the Czech Republic, Czech Republic;
The Danish Council for Independent Research | Natural Sciences, the VILLUM FONDEN and Danish National Research Foundation (DNRF), Denmark;
Helsinki Institute of Physics (HIP), Finland;
Commissariat \`{a} l'Energie Atomique (CEA) and Institut National de Physique Nucl\'{e}aire et de Physique des Particules (IN2P3) and Centre National de la Recherche Scientifique (CNRS), France;
Bundesministerium f\"{u}r Forschung, Technologie und Raumfahrt (BMFTR) and GSI Helmholtzzentrum f\"{u}r Schwerionenforschung GmbH, Germany;
General Secretariat for Research and Technology, Ministry of Education, Research and Religions, Greece;
National Research, Development and Innovation Office, Hungary;
Department of Atomic Energy Government of India (DAE), Department of Science and Technology, Government of India (DST), University Grants Commission, Government of India (UGC) and Council of Scientific and Industrial Research (CSIR), India;
National Research and Innovation Agency - BRIN, Indonesia;
Istituto Nazionale di Fisica Nucleare (INFN), Italy;
Japanese Ministry of Education, Culture, Sports, Science and Technology (MEXT) and Japan Society for the Promotion of Science (JSPS) KAKENHI, Japan;
Consejo Nacional de Ciencia (CONACYT) y Tecnolog\'{i}a, through Fondo de Cooperaci\'{o}n Internacional en Ciencia y Tecnolog\'{i}a (FONCICYT) and Direcci\'{o}n General de Asuntos del Personal Academico (DGAPA), Mexico;
Nederlandse Organisatie voor Wetenschappelijk Onderzoek (NWO), Netherlands;
The Research Council of Norway, Norway;
Pontificia Universidad Cat\'{o}lica del Per\'{u}, Peru;
Ministry of Science and Higher Education, National Science Centre and WUT ID-UB, Poland;
Korea Institute of Science and Technology Information and National Research Foundation of Korea (NRF), Republic of Korea;
Ministry of Education and Scientific Research, Institute of Atomic Physics, Ministry of Research and Innovation and Institute of Atomic Physics and Universitatea Nationala de Stiinta si Tehnologie Politehnica Bucuresti, Romania;
Ministerstvo skolstva, vyskumu, vyvoja a mladeze SR, Slovakia;
National Research Foundation of South Africa, South Africa;
Swedish Research Council (VR) and Knut \& Alice Wallenberg Foundation (KAW), Sweden;
European Organization for Nuclear Research, Switzerland;
Suranaree University of Technology (SUT), National Science and Technology Development Agency (NSTDA) and National Science, Research and Innovation Fund (NSRF via PMU-B B05F650021), Thailand;
Turkish Energy, Nuclear and Mineral Research Agency (TENMAK), Turkey;
National Academy of  Sciences of Ukraine, Ukraine;
Science and Technology Facilities Council (STFC), United Kingdom;
National Science Foundation of the United States of America (NSF) and United States Department of Energy, Office of Nuclear Physics (DOE NP), United States of America.
In addition, individual groups or members have received support from:
Czech Science Foundation (grant no. 23-07499S), Czech Republic;
FORTE project, reg.\ no.\ CZ.02.01.01/00/22\_008/0004632, Czech Republic, co-funded by the European Union, Czech Republic;
European Research Council (grant no. 950692), European Union;
Deutsche Forschungs Gemeinschaft (DFG, German Research Foundation) ``Neutrinos and Dark Matter in Astro- and Particle Physics'' (grant no. SFB 1258), Germany;
ICSC - National Research Center for High Performance Computing, Big Data and Quantum Computing and FAIR - Future Artificial Intelligence Research, funded by the NextGenerationEU program (Italy).

\end{acknowledgement}

\bibliographystyle{utphys}
\bibliography{paper.bib}

\newpage
\appendix

%
%

\section{The ALICE Collaboration}
\label{app:collab}
\begin{flushleft} 
\small

I.J.~Abualrob\,\orcidlink{0009-0005-3519-5631}\,$^{\rm 114}$, 
S.~Acharya\,\orcidlink{0000-0002-9213-5329}\,$^{\rm 50}$, 
G.~Aglieri Rinella\,\orcidlink{0000-0002-9611-3696}\,$^{\rm 32}$, 
L.~Aglietta\,\orcidlink{0009-0003-0763-6802}\,$^{\rm 24}$, 
M.~Agnello\,\orcidlink{0000-0002-0760-5075}\,$^{\rm 29}$, 
N.~Agrawal\,\orcidlink{0000-0003-0348-9836}\,$^{\rm 25}$, 
Z.~Ahammed\,\orcidlink{0000-0001-5241-7412}\,$^{\rm 133}$, 
S.~Ahmad\,\orcidlink{0000-0003-0497-5705}\,$^{\rm 15}$, 
I.~Ahuja\,\orcidlink{0000-0002-4417-1392}\,$^{\rm 36}$, 
ZUL.~Akbar$^{\rm 81}$, 
A.~Akindinov\,\orcidlink{0000-0002-7388-3022}\,$^{\rm 139}$
% ...

\section*{Affiliation Notes}

$^{\rm I}$ Deceased\\
$^{\rm II}$ Also at: Max-Planck-Institut fur Physik, Munich, Germany\\
$^{\rm III}$ Also at: Czech Technical University in Prague (CZ)\\
$^{\rm IV}$ Also at: Italian National Agency for New Technologies, Energy and Sustainable Economic Development (ENEA), Bologna, Italy\\
$^{\rm V}$ Also at: Instituto de Fisica da Universidade de Sao Paulo\\
$^{\rm VI}$ Also at: Dipartimento DET del Politecnico di Torino, Turin, Italy\\
$^{\rm VII}$ Also at: Department of Applied Physics, Aligarh Muslim University, Aligarh, India\\
$^{\rm VIII}$ Also at: Institute of Theoretical Physics, University of Wroclaw, Poland\\
$^{\rm IX}$ Also at: Facultad de Ciencias, Universidad Nacional Aut\'{o}noma de M\'{e}xico, Mexico City, Mexico\\

\section*{Collaboration Institutes}

$^{1}$ A.I. Alikhanyan National Science Laboratory (Yerevan Physics Institute) Foundation, Yerevan, Armenia\\
$^{2}$ AGH University of Krakow, Cracow, Poland\\
$^{3}$ Bogolyubov Institute for Theoretical Physics, National Academy of Sciences of Ukraine, Kyiv, Ukraine\\
$^{4}$ Bose Institute, Department of Physics  and Centre for Astroparticle Physics and Space Science (CAPSS), Kolkata, India\\
$^{5}$ California Polytechnic State University, San Luis Obispo, California, United States\\
$^{6}$ Central China Normal University, Wuhan, China\\
$^{7}$ Centro de Aplicaciones Tecnol\'{o}gicas y Desarrollo Nuclear (CEADEN), Havana, Cuba\\
$^{8}$ Centro de Investigaci\'{o}n y de Estudios Avanzados (CINVESTAV), Mexico City and M\'{e}rida, Mexico\\
$^{9}$ Chicago State University, Chicago, Illinois, United States\\
$^{10}$ China Nuclear Data Center, China Institute of Atomic Energy, Beijing, China\\
$^{11}$ China University of Geosciences, Wuhan, China\\
$^{12}$ Chungbuk National University, Cheongju, Republic of Korea\\
$^{13}$ Comenius University Bratislava, Faculty of Mathematics, Physics and Informatics, Bratislava, Slovak Republic\\
$^{14}$ Creighton University, Omaha, Nebraska, United States\\
$^{15}$ Department of Physics, Aligarh Muslim University, Aligarh, India\\
$^{16}$ Department of Physics, Pusan National University, Pusan, Republic of Korea\\
$^{17}$ Department of Physics, Sejong University, Seoul, Republic of Korea\\
$^{18}$ Department of Physics, University of California, Berkeley, California, United States\\
$^{19}$ Department of Physics, University of Oslo, Oslo, Norway\\
$^{20}$ Department of Physics and Technology, University of Bergen, Bergen, Norway\\
$^{21}$ Dipartimento di Fisica, Universit\`{a} di Pavia, Pavia, Italy\\
$^{22}$ Dipartimento di Fisica dell'Universit\`{a} and Sezione INFN, Cagliari, Italy\\
$^{23}$ Dipartimento di Fisica dell'Universit\`{a} and Sezione INFN, Trieste, Italy\\
$^{24}$ Dipartimento di Fisica dell'Universit\`{a} and Sezione INFN, Turin, Italy\\
$^{25}$ Dipartimento di Fisica e Astronomia dell'Universit\`{a} and Sezione INFN, Bologna, Italy\\
$^{26}$ Dipartimento di Fisica e Astronomia dell'Universit\`{a} and Sezione INFN, Catania, Italy\\
$^{27}$ Dipartimento di Fisica e Astronomia dell'Universit\`{a} and Sezione INFN, Padova, Italy\\
$^{28}$ Dipartimento di Fisica `E.R.~Caianiello' dell'Universit\`{a} and Gruppo Collegato INFN, Salerno, Italy\\
$^{29}$ Dipartimento DISAT del Politecnico and Sezione INFN, Turin, Italy\\
$^{30}$ Dipartimento di Scienze MIFT, Universit\`{a} di Messina, Messina, Italy\\
$^{31}$ Dipartimento Interateneo di Fisica `M.~Merlin' and Sezione INFN, Bari, Italy\\
$^{32}$ European Organization for Nuclear Research (CERN), Geneva, Switzerland\\
$^{33}$ Faculty of Electrical Engineering, Mechanical Engineering and Naval Architecture, University of Split, Split, Croatia\\
$^{34}$ Faculty of Nuclear Sciences and Physical Engineering, Czech Technical University in Prague, Prague, Czech Republic\\
$^{35}$ Faculty of Physics, Sofia University, Sofia, Bulgaria\\
$^{36}$ Faculty of Science, P.J.~\v{S}af\'{a}rik University, Ko\v{s}ice, Slovak Republic\\
$^{37}$ Faculty of Technology, Environmental and Social Sciences, Bergen, Norway\\
$^{38}$ Frankfurt Institute for Advanced Studies, Johann Wolfgang Goethe-Universit\"{a}t Frankfurt, Frankfurt, Germany\\
$^{39}$ Fudan University, Shanghai, China\\
$^{40}$ Gangneung-Wonju National University, Gangneung, Republic of Korea\\
$^{41}$ Gauhati University, Department of Physics, Guwahati, India\\
$^{42}$ Helmholtz-Institut f\"{u}r Strahlen- und Kernphysik, Rheinische Friedrich-Wilhelms-Universit\"{a}t Bonn, Bonn, Germany\\
$^{43}$ Helsinki Institute of Physics (HIP), Helsinki, Finland\\
$^{44}$ High Energy Physics Group,  Universidad Aut\'{o}noma de Puebla, Puebla, Mexico\\
$^{45}$ Horia Hulubei National Institute of Physics and Nuclear Engineering, Bucharest, Romania\\
$^{46}$ HUN-REN Wigner Research Centre for Physics, Budapest, Hungary\\
$^{47}$ Indian Institute of Technology Bombay (IIT), Mumbai, India\\
$^{48}$ Indian Institute of Technology Indore, Indore, India\\
$^{49}$ INFN, Laboratori Nazionali di Frascati, Frascati, Italy\\
$^{50}$ INFN, Sezione di Bari, Bari, Italy\\
$^{51}$ INFN, Sezione di Bologna, Bologna, Italy\\
$^{52}$ INFN, Sezione di Cagliari, Cagliari, Italy\\
$^{53}$ INFN, Sezione di Catania, Catania, Italy\\
$^{54}$ INFN, Sezione di Padova, Padova, Italy\\
$^{55}$ INFN, Sezione di Pavia, Pavia, Italy\\
$^{56}$ INFN, Sezione di Torino, Turin, Italy\\
$^{57}$ INFN, Sezione di Trieste, Trieste, Italy\\
$^{58}$ Inha University, Incheon, Republic of Korea\\
$^{59}$ Institute for Gravitational and Subatomic Physics (GRASP), Utrecht University/Nikhef, Utrecht, Netherlands\\
$^{60}$ Institute of Experimental Physics, Slovak Academy of Sciences, Ko\v{s}ice, Slovak Republic\\
$^{61}$ Institute of Physics, Homi Bhabha National Institute, Bhubaneswar, India\\
$^{62}$ Institute of Physics of the Czech Academy of Sciences, Prague, Czech Republic\\
$^{63}$ Institute of Space Science (ISS), Bucharest, Romania\\
$^{64}$ Institut f\"{u}r Kernphysik, Johann Wolfgang Goethe-Universit\"{a}t Frankfurt, Frankfurt, Germany\\
$^{65}$ Instituto de Ciencias Nucleares, Universidad Nacional Aut\'{o}noma de M\'{e}xico, Mexico City, Mexico\\
$^{66}$ Instituto de F\'{i}sica, Universidade Federal do Rio Grande do Sul (UFRGS), Porto Alegre, Brazil\\
$^{67}$ Instituto de F\'{\i}sica, Universidad Nacional Aut\'{o}noma de M\'{e}xico, Mexico City, Mexico\\
$^{68}$ iThemba LABS, National Research Foundation, Somerset West, South Africa\\
$^{69}$ Jeonbuk National University, Jeonju, Republic of Korea\\
$^{70}$ Johann-Wolfgang-Goethe Universit\"{a}t Frankfurt Institut f\"{u}r Informatik, Fachbereich Informatik und Mathematik, Frankfurt, Germany\\
$^{71}$ Korea Institute of Science and Technology Information, Daejeon, Republic of Korea\\
$^{72}$ Laboratoire de Physique Subatomique et de Cosmologie, Universit\'{e} Grenoble-Alpes, CNRS-IN2P3, Grenoble, France\\
$^{73}$ Lawrence Berkeley National Laboratory, Berkeley, California, United States\\
$^{74}$ Lund University Department of Physics, Division of Particle Physics, Lund, Sweden\\
$^{75}$ Nagasaki Institute of Applied Science, Nagasaki, Japan\\
$^{76}$ Nara Women{'}s University (NWU), Nara, Japan\\
$^{77}$ National and Kapodistrian University of Athens, School of Science, Department of Physics , Athens, Greece\\
$^{78}$ National Centre for Nuclear Research, Warsaw, Poland\\
$^{79}$ National Institute of Science Education and Research, Homi Bhabha National Institute, Jatni, India\\
$^{80}$ National Nuclear Research Center, Baku, Azerbaijan\\
$^{81}$ National Research and Innovation Agency - BRIN, Jakarta, Indonesia\\
$^{82}$ Niels Bohr Institute, University of Copenhagen, Copenhagen, Denmark\\
$^{83}$ Nikhef, National institute for subatomic physics, Amsterdam, Netherlands\\
$^{84}$ Nuclear Physics Group, STFC Daresbury Laboratory, Daresbury, United Kingdom\\
$^{85}$ Nuclear Physics Institute of the Czech Academy of Sciences, Husinec-\v{R}e\v{z}, Czech Republic\\
$^{86}$ Oak Ridge National Laboratory, Oak Ridge, Tennessee, United States\\
$^{87}$ Ohio State University, Columbus, Ohio, United States\\
$^{88}$ Physics department, Faculty of science, University of Zagreb, Zagreb, Croatia\\
$^{89}$ Physics Department, Panjab University, Chandigarh, India\\
$^{90}$ Physics Department, University of Jammu, Jammu, India\\
$^{91}$ Physics Program and International Institute for Sustainability with Knotted Chiral Meta Matter (WPI-SKCM$^{2}$), Hiroshima University, Hiroshima, Japan\\
$^{92}$ Physikalisches Institut, Eberhard-Karls-Universit\"{a}t T\"{u}bingen, T\"{u}bingen, Germany\\
$^{93}$ Physikalisches Institut, Ruprecht-Karls-Universit\"{a}t Heidelberg, Heidelberg, Germany\\
$^{94}$ Physik Department, Technische Universit\"{a}t M\"{u}nchen, Munich, Germany\\
$^{95}$ Politecnico di Bari and Sezione INFN, Bari, Italy\\
$^{96}$ Research Division and ExtreMe Matter Institute EMMI, GSI Helmholtzzentrum f\"ur Schwerionenforschung GmbH, Darmstadt, Germany\\
$^{97}$ Saga University, Saga, Japan\\
$^{98}$ Saha Institute of Nuclear Physics, Homi Bhabha National Institute, Kolkata, India\\
$^{99}$ School of Physics and Astronomy, University of Birmingham, Birmingham, United Kingdom\\
$^{100}$ Secci\'{o}n F\'{\i}sica, Departamento de Ciencias, Pontificia Universidad Cat\'{o}lica del Per\'{u}, Lima, Peru\\
$^{101}$ Stefan Meyer Institut f\"{u}r Subatomare Physik (SMI), Vienna, Austria\\
$^{102}$ SUBATECH, IMT Atlantique, Nantes Universit\'{e}, CNRS-IN2P3, Nantes, France\\
$^{103}$ Sungkyunkwan University, Suwon City, Republic of Korea\\
$^{104}$ Suranaree University of Technology, Nakhon Ratchasima, Thailand\\
$^{105}$ Technical University of Ko\v{s}ice, Ko\v{s}ice, Slovak Republic\\
$^{106}$ The Henryk Niewodniczanski Institute of Nuclear Physics, Polish Academy of Sciences, Cracow, Poland\\
$^{107}$ The University of Texas at Austin, Austin, Texas, United States\\
$^{108}$ Universidad Aut\'{o}noma de Sinaloa, Culiac\'{a}n, Mexico\\
$^{109}$ Universidade de S\~{a}o Paulo (USP), S\~{a}o Paulo, Brazil\\
$^{110}$ Universidade Estadual de Campinas (UNICAMP), Campinas, Brazil\\
$^{111}$ Universidade Federal do ABC, Santo Andre, Brazil\\
$^{112}$ Universitatea Nationala de Stiinta si Tehnologie Politehnica Bucuresti, Bucharest, Romania\\
$^{113}$ University of Derby, Derby, United Kingdom\\
$^{114}$ University of Houston, Houston, Texas, United States\\
$^{115}$ University of Jyv\"{a}skyl\"{a}, Jyv\"{a}skyl\"{a}, Finland\\
$^{116}$ University of Kansas, Lawrence, Kansas, United States\\
$^{117}$ University of Liverpool, Liverpool, United Kingdom\\
$^{118}$ University of Science and Technology of China, Hefei, China\\
$^{119}$ University of South-Eastern Norway, Kongsberg, Norway\\
$^{120}$ University of Tennessee, Knoxville, Tennessee, United States\\
$^{121}$ University of the Witwatersrand, Johannesburg, South Africa\\
$^{122}$ University of Tokyo, Tokyo, Japan\\
$^{123}$ University of Tsukuba, Tsukuba, Japan\\
$^{124}$ Universit\"{a}t M\"{u}nster, Institut f\"{u}r Kernphysik, M\"{u}nster, Germany\\
$^{125}$ Universit\'{e} Clermont Auvergne, CNRS/IN2P3, LPC, Clermont-Ferrand, France\\
$^{126}$ Universit\'{e} de Lyon, CNRS/IN2P3, Institut de Physique des 2 Infinis de Lyon, Lyon, France\\
$^{127}$ Universit\'{e} de Strasbourg, CNRS, IPHC UMR 7178, F-67000 Strasbourg, France, Strasbourg, France\\
$^{128}$ Universit\'{e} Paris-Saclay, Centre d'Etudes de Saclay (CEA), IRFU, D\'{e}partment de Physique Nucl\'{e}aire (DPhN), Saclay, France\\
$^{129}$ Universit\'{e}  Paris-Saclay, CNRS/IN2P3, IJCLab, Orsay, France\\
$^{130}$ Universit\`{a} degli Studi di Foggia, Foggia, Italy\\
$^{131}$ Universit\`{a} del Piemonte Orientale, Vercelli, Italy\\
$^{132}$ Universit\`{a} di Brescia, Brescia, Italy\\
$^{133}$ Variable Energy Cyclotron Centre, Homi Bhabha National Institute, Kolkata, India\\
$^{134}$ Warsaw University of Technology, Warsaw, Poland\\
$^{135}$ Wayne State University, Detroit, Michigan, United States\\
$^{136}$ Yale University, New Haven, Connecticut, United States\\
$^{137}$ Yildiz Technical University, Istanbul, Turkey\\
$^{138}$ Yonsei University, Seoul, Republic of Korea\\
$^{139}$ Affiliated with an institute formerly covered by a cooperation agreement with CERN\\
$^{140}$ Affiliated with an international laboratory covered by a cooperation agreement with CERN.\\

\end{flushleft} 
 
\end{document}